\documentclass[aps,prl,preprint,nopacs,superscriptaddress]{revtex4}

\usepackage{amsmath,cleveref,amssymb,amsfonts,graphicx,multirow,color,bm,textcomp,mathtools}

\newcommand{\bee}{\begin{eqnarray}}
\newcommand{\ee}{\end{eqnarray}}
\newcommand{\bma}{\begin{pmatrix}}
\newcommand{\ema}{\end{pmatrix}}
\newcommand{\balig}{\begin{align}}
\newcommand{\ealig}{\end{align}}

\newcommand{\bZ}{\mathbb{Z}}
\newcommand{\ba}{\begin{align}}
\newcommand{\ea}{\end{align}}

\newcommand{\ignore}[1]{}

\begin{document}

\title{Chiral Majorana Fermion Modes on the Surface of Superconducting Topological Insulators}

%
\author{Ching-Kai Chiu}
\affiliation{Condensed Matter Theory Center and Joint Quantum Institute and Station Q Maryland, Department of Physics, University of Maryland, College Park, MD 20742, USA}

\author{Guang~Bian}
\affiliation {Department of Physics and Astronomy, University of Missouri, Columbia, Missouri 65211, USA}

\author{Hao~Zheng}
\affiliation {Laboratory for Topological Quantum Matter and Spectroscopy (B7), Department of Physics, Princeton University, Princeton, New Jersey 08544, USA}

\author{Jiaxin~Yin}
\affiliation {Laboratory for Topological Quantum Matter and Spectroscopy (B7), Department of Physics, Princeton University, Princeton, New Jersey 08544, USA}

\author{Songtian~S.~Zhang}
\affiliation {Laboratory for Topological Quantum Matter and Spectroscopy (B7), Department of Physics, Princeton University, Princeton, New Jersey 08544, USA}

\author{Su-Yang~Xu}
\affiliation {Laboratory for Topological Quantum Matter and Spectroscopy (B7), Department of Physics, Princeton University, Princeton, New Jersey 08544, USA}

\author{M. Zahid~Hasan}
\affiliation {Laboratory for Topological Quantum Matter and Spectroscopy (B7), Department of Physics, Princeton University, Princeton, New Jersey 08544, USA}
\affiliation {Lawrence Berkeley National Laboratory, Berkeley, CA 94720, USA}

%
%
%
%

%
%
%
%
%

%


\begin{abstract}

 The surface of superconducting topological insulators (STIs) has been recognized as an effective $p\pm ip$ superconductivity platform for realizing elusive Majorana fermions. Chiral Majorana modes (CMMs), which are different from Majorana bound states localized at points, can be achieved readily in experiments by depositing a ferromagnetic overlayer on top of the STI surface. Here we simulate this heterostructure by employing a realistic tight-binding model and show that the CMM appears on the edge of the ferromagnetic islands only after the superconducting gap is inverted by the exchange coupling between the ferromagnet and the STI. In addition, multiple CMMs can be generated by tuning the chemical potential of the topological insulator. These results can be applied to both proximity-effect induced superconductivity in topological insulators and intrinsic STI compounds such as PbTaSe$_2$, BiPd and their chemical analogues, providing a route to engineering CMMs in those materials.

\end{abstract}

\maketitle

The search for elusive Majorana fermions in condensed matter systems has been under intense study for potential applications of Majorana bound states (MBSs) with exotic braiding properties in fault-tolerant topological computations \cite{Alicea, Nayak}. Despite strong experimental evidence of MBSs in nanowires and atomic chains, stringent requirements on experimental conditions hinder a conclusive identification of MBSs in these quasi-1D systems \cite{Kitaev, Mourik, Das, Deng, Yazdani, Marcus,Marcus2}. On the other hand, the interplay of topological insulators (TIs) and superconductivity creates intriguing circumstances where an effective time-reversal symmetric $p\pm ip$ superconductivity can naturally emerge as a consequence of the spin-momentum-locking of topological surface quasiparticles. Many materials exhibiting both nontrivial band topology and superconductivity have been experimentally discovered and synthesized \cite{TI1, TI2, Review, Chiu_mod, Andreas, PbTaSe2, BiPd, Chang167, FeSeML, CuBiSe, TS-dop1, TS-dop2, TS-dop3, TlBiTe, Yacoby, BiSeHetero, BiSeHTc, DasSarma}. Surface $p\pm ip$ superconductivity can be straightforwardly tested in those compounds, enabling a solid-state environment to realize the Kitaev picture of Majorana fermions. 
If one further breaks the time reversal symmetry in these systems, MBSs can be found at the boundaries and defects of the surface exhibiting topological superconductivity. Significant experimental progress \cite{Jia, KangWang} has revealed the potential existence of two distinct MBSs; a zero-energy MBS localized on the end of the vortex, and a CMM with \emph{linear} energy dispersion at the edge, the two of which are entirely different entities. Although the existence of CMMs has been claimed in superconducting Chern insulator films \cite{KangWang}, studies of CMMs are rare in the literature~\cite{Fu_CMM,Akhmerov_CMM,Jay_CMM}.  
Here, we propose a different approach for realizing CMMs by depositing a ferromagnetic layer (FI) on top of a  STI surface. We find that CMMs can be visualized at the edge of the ferromagnetic islands if the magnetic exchange coupling with the underlying TI surface is properly tuned. The chirality of the CMMs depends on the magnetization direction of the ferromagnetic domains. The integration of multiple ferromagnetic islands on a single surface allows a convenient and accessible platform for probing the rich physics arising from interactions between CMMs. 

To reveal the topology of the STI surface, we first focus on the low-energy BdG Hamiltonian of a Dirac cone surface with s-wave superconductivity 
\bee
\hat{H}_{\rm{BdG}}(k)= C_k^\dagger
\bma 
0 & \nu(k_x-ik_y) & 0 & -\Delta \\
\nu(k_x +ik_y) & 0 & \Delta & 0 \\
0 & \Delta & 0  & \nu(k_x+ik_y) \\
-\Delta & 0 & \nu(k_x-ik_y) & 0 \\
\ema
C_k,
\ee  
where $\Delta$ is the superconducting order parameter, $\nu$ is the Fermi velocity of the surface cone, and the Fermion operator $C_k=(c_{\uparrow k}, c_{\downarrow k}, c_{\uparrow -k}^\dagger, c_{\downarrow -k}^\dagger)^T$. By transforming the Fermion basis to $C_k'=\frac{1}{\sqrt{2}}(c_{\uparrow k}-c_{\downarrow -k}^\dagger,c_{\downarrow k}-c_{\uparrow -k}^\dagger,c_{\downarrow k}+c_{\uparrow -k}^\dagger,c_{\uparrow k}+c_{\downarrow -k}^\dagger)$ the Hamiltonian can be further block-diagonalized 
\bee
\hat{H}_{\rm{BdG}}(k)= C_k'^\dagger
\bma 
\Delta & \nu(k_x-ik_y) & 0 & 0 \\
\nu(k_x +ik_y) & -\Delta & 0 & 0 \\
0 & 0 & \Delta  & \nu(k_x+ik_y) \\
0 & 0 & \nu(k_x-ik_y) & -\Delta \\
\ema
C_k',
\ee  
The quadratic Hamiltonian can be rewritten in an economical matter $h(k)=\Delta \sigma_z \otimes \tau_0 +\nu k_x \sigma_x \otimes \tau_0 + \nu k_y \sigma_y \otimes \tau_z $. Each block preserves its own particle-hole symmetry with the symmetry operator $C=\mp \sigma_x K$, where $K$ is the complex conjugate operator; hence, the first (second) block is an effective $p\pm ip$ superconducting Hamiltonian. From comparison with physical $p+ip$ superconductivity, the off-diagonal terms stem from the surface Dirac cone, instead of the $p$-wave superconductor order parameter. Furthermore, these two blocks are time-reversal partners connected by the time-reversal symmetry operator $T=i\sigma_z \tau_yK$. Although the superconducting surface possesses non-trivial topology \cite{Nontrivial_surface_chiu}, gapless bound states are absent since the surface of the bulk has no boundary. In addition, in order to observe CMMs at the boundary of this 2D superconductor, one has to break the time reversal symmetry. Instead of applying an external magnetic field as proposed by Fu and Kane~\cite{FuKane_SC_STI, TS-9}, we adopt a different approach by introducing a ferromagnetic insulator layer coupled to the STI surface as schematically shown in Fig.~1(a).  The FI generates the Zeeman field and we specifically choose the splitting to be along the $z$ direction $M_z\sigma_z\otimes \tau_z$ in the new basis. Consider $M_z>\Delta>0$ as $x>0$ and $M_z=0,\ \Delta> 0$ as $x<0$ so that the ill-defined $k_x$ momentum leads to $k_x\rightarrow -i\partial/\partial x$. By solving the eigenvalue problem of $h(k)+M_z\sigma_z\otimes \tau_z$, the chiral Majorana state 
\bee
\Phi_{\rm{CMM}}=\frac{e^{-(M_z-\Delta)x/\nu}}{\sqrt{2}}(0,0,e^{i\pi/4},e^{-i\pi/4}) \label{CMM_wf}
\ee
is localized at $x=0$ with linear dispersion $-\nu k_y$. Hence, the boundary of the FI can host a CMM. This proposed structure can be readily fabricated by modern molecular beam epitaxy technique. Here, we study the time-reversal breaking effect of a FI on the STI surface by using a tight-binding description of the system. 

In order to show the existence of the CMM as predicted by the low energy model, we introduce in the tight-binding model a FI layer on the top of a STI. The BdG Hamiltonian can be written as
\begin{subequations} \label{HFK}
\bee
\hat{H}(k_x,k_y)=\hat{H}_{\rm{TISC}}+\hat{H}_{\rm{M}}+\hat{H}_T,
\ee
where 
\begin{small}
\begin{align}
\hat{H}_{\rm{TISC}}=&\sum_{0 < z \leq L_{\rm{TI}}} \Pi_z^\dagger \Big \{ \big [ M -2B_1 - 2B_2 (2- c(k_x,k_y)) \big ] \tau_z \sigma_z s_0 \nonumber \\
+& A_2 \sin ak_x  \tau_0 \sigma_x s_x + A_2 \sin ak_y \tau_z \sigma_x s_y - \mu \tau_z \sigma_0 s_0+\Delta\tau_y \sigma_0 s_y  \Big \} \Pi_z \nonumber \\
+&\sum_{0 < z < L_{\rm{TI}}} \big [ \Pi_{z+a}^\dagger (B_1\tau_z \sigma_z s_0 - \frac{iA_1}{2}\tau_0 \sigma_x s_z)\Pi_z + h.c. \big ], \label{TI_FK} \\
\hat{H}_{\rm{FI}}=& \Pi^\dagger_{0} \big [ M_x \tau_z \sigma_0 s_x + M_y \tau_0 \sigma_0 s_y + M_z \tau_z \sigma_0 s_z  \big ] \Pi_{0},   \\
\hat{H}_T=&T\Pi^\dagger_{0} \tau_z \sigma_0 s_0 \Pi_{a} + h.c.,
\end{align}
\end{small}
\end{subequations}
\noindent where $c(k_x,k_y)=\cos ak_x + \cos ak_y$, by choosing the lattice constant $a=3\mathrm~{\mathrm{\AA}}$, the location label $z\in \bZ \cdot 3~\mathrm{\AA}$, and the thickness of the TI $L_{\rm{TI}}=90~\mathrm{\AA}$. The Hamiltonian is expressed in the Nambu basis of fermion annihilation and creation operators $\Pi=(b_{\uparrow}\ b_{\downarrow}\ d_{\uparrow}\ d_{\downarrow}\ b_{\uparrow}^\dagger\ b_{\downarrow}^\dagger\ d_{\uparrow}^\dagger\ d_{\downarrow}^\dagger)^T$, where $\tau_\alpha,\ \sigma_\beta,\ $and $s_\gamma$ describe particle-hole, orbital, and spin degrees of freedom respectively. The FI $\hat{H}_{\rm{FI}}$ is simply described by the Zeeman effect only along the $z$ direction and  $M_z=50$~meV is a reasonable estimation for the Zeeman splitting at the Dirac point of TI surface states~\cite{Chen659, Potter, TS-8}. The presence of the coupling $\hat{H}_T$ between the FI and the surface of the STI is the main difference between the tight-binding model and the low-energy model. Due to the experimental realization of STI in Bi$_2$Se$_3$~\cite{Jia_STI}, the parameters used in the STI Hamiltonian $\hat{H}_{\rm{TISC}}$ are mainly based on its properties~\cite{Zhang:2009aa};  in units of eV: bulk gap $M=0.28$, chemical potential $\mu=0.05$, $A_1=2.2/a$, $A_2=4.1/a$, $B_1=10/a^2$, $B_2=56.5/a^2$ (unless specified otherwise). The band structure of this topological insulator is shown in Fig.~1(b). We first intentionally choose the superconducting order parameter to be $\Delta=20$~meV, which is comparable to the TI parameters; a more realistic value for $\Delta$ will be considered later. In the following, we compute several lowest energy states and projected edge spectrum of the FI to determine if the interface between the STI and the FI can be a platform for the non-trivial topology. 

	We first examine the topological phase transition at the boundless FI/STI interface by varying the coupling $T$ between the FI overlayer and the STI surface. When the coupling is zero ($T=0$), the spectrum in Fig.~2(a) clearly shows a $50$~meV ferromagnetic gap in the FI layer and a $20$~meV superconducting gap, which gaps out the surface Dirac surface cone. As the coupling $T$ grows, the surface gap becomes smaller. At $T=0.122$~eV, as shown in Fig.~1(c) and Fig.~2(b) the bulk gap closes and this linear dispersion band touching indicates that the Chern number changes by one; hence, the interface is expected to possess non-zero Chern number $C=1$ as $T>0.122$~eV. If there is an open boundary condition, a CMM is expected to be present at the edge of the interface. Thus, we transfer the momentum $k_y$ to real space $y$ in the Hamiltonian (\ref{HFK}) and choose the width of a ribbon-shaped FI island to be $L=100a$. We compute the density of states at one edge of the FI for three different couplings $T$. As shown in the projected edge spectrum Fig.~2(f), when the coupling is large with $T=0.2$~eV, a single CMM appears on the edge. However, here we have created an ideal situation where we have used a large value for the superconducting order parameter and have adjusted the chemical potential in the bulk gap and close to the surface Dirac point.

	Now, we consider more realistic situations. We first adjust the chemical potential to be $0.35$~eV greater than the bulk gap of the TI, since the TI is expected to be metallic in order for superconductivity to be induced. In some TI/SC heterostructures, the Fermi level was also found to be in the bulk band energy region \cite{BiSeHetero, Jia}.  However, there is generally no such constraint on systems with proximity-induced superconductivity~\cite{Chiu_proximity}.  As the coupling $T$ increases from 0, the situation is more complicated than simply crossing a gap-closing point. Since there are multiple band touchings which persist for moderate values of $T$,  as shown in Fig.~3(a), it is difficult to determine the change of the topology after the transition. Fortunately, at large $T$, the interface can still be gapped, and hosts two CMMs at the edges, see Fig.~3(b).  Therefore, the interface is still in a topologically non-trivial phase. Since two CMMs are equivalent to a single chiral electron mode, the interface becomes an effective Chern insulator \cite{Chang167}.

	We now consider a more realistic value for the superconducting order parameter, $\Delta=3$~meV, and further adjust the chemical potential $\mu=0.01$~eV to be close to the surface Dirac point. As the decay length of the Majorana wavefunctions dramatically increase due to the small order parameter, the CMMs on the two opposite edges overlap in real space, hybridize, and are gapped in energy for a FI ribbon with a width of 100a. The hybridization gap exhibits exponential decay with an oscillation as the width (L) of the FI ribbon increases, as shown in Fig. 3(c). Although the oscillation is not expected in the low-energy model (\ref{CMM_wf}), the behavior of hybridization gap depends on the details of the system~\cite{Marcus, Chiu_proximity}. In order to have gapless CMMs, we specifically choose a large width for the FI ribbon $L=2000a \sim 0.6\mu m$, a size which can be reasonably acheived in experiment. The projected edge spectrum (Fig.~3(d)) shows a $0.043$meV gap, which is relatively small compared to the superconducting gap $3$~meV. On the other hand, the out-of-plane magnetization is the crucial ingredient for the realization of Majorana modes, as a Zeeman effect along the normal direction of the TI surface is required to open a gap at the Dirac point of the topological surface states. However, the magnetization favors the in-plane direction in some FI films \cite{Jiaxin}; in order to circumvent this problem, a weak magnetic field can be applied in the $z$ direction and generate a nonzero out-of-plane component in the magnetization. We consider two Zeeman-effect directions $M_x=M_z=0.05/\sqrt{2}$ eV and $M_y=M_z=0.05/\sqrt{2}$ eV, both of which include in-plane and out-of-plane components. Fortunately, the presence of the in-plane does not destroy the CMMs. As shown in Fig.~3(e,f), the $x$ component of the magnetic field (parallel to the ribbon edge) does not alter the spectrum, and in $y$ (perpendicular to the ribbon edge), only the energy dispersion of the in-gap states changes, while the CMM remains intact. 
	
\section{Discussions}

In TI/superconductor heterostructures, the Fermi level can lie in the bulk band gap if the TI is bulk insulating, for example, the compound BiSbTeSe$_2$ \cite{YongChen, Madhab}.  This is due to the fact that the surface superconductivity of topological insulators is induced only by the proximity effect from the superconductor attached to the TI film. As we demonstrated in the model simulation, in the bulk insulating case, the Majorana states come entirely from the Dirac surface states of the TI, and thus, one can obtain a CMM on the boundary of each magnetic island by continuously tuning the exchange coupling between the TI and the ferromagnetic overlayer. On the other hand, intrinsic STIs have to possess a finite bulk Fermi surface in order to have the Cooper instability of the Fermi surface and thus induce superconductivity. In such cases, the bulk carriers also participate in Cooper pairing and contribute to the Majorana spectrum. This complication leads to either no or multiple CMMs, depending on the details of the magnetic coupling. STIs can be classified into two types. The first is chemically doped TIs such as Cu$_x$Bi$_2$Se$_3$ and Tl$_x$Bi$_2$Te$_3$ \cite{CuBiSe, TlBiTe}. This class of materials has a simple bulk Fermi surface, as the bulk pockets are usually from the parabolic conduction band of the TI compound.  However, the chemical doping inevitably introduces uncontrollable disorders into the system. The other class are materials such as stoichiometric superconducting PbTaSe$_2$ and BiPd \cite{PbTaSe2, BiPd}. While there is no chemical disorder in these compounds, the bulk fermi surface consists of multiple bulk pockets, so that it is more complicated than those of conventional TIs such as Bi$_2$Se$_3$. Nevertheless, these materials, regardless of being single compounds or heterostructures, or stoichiometric or chemically doped materials, all potentially host nontrivial surface superconductivity and provide multiple avenues through which the implementation of CMMs with this hybrid-film based platform can be pursued.

As realistic values for some of the experimental parameters are unknown, such as the coupling $T$, it is difficult to precisely determine the topology of this heterostructure. Our result shows that with the proper parameters, the CMMs can be realized on the edges of the FI and that the parameters for which there exists a topologically nontrivial region can be narrowed down for future experiment. We provide an experimental recipe for confirming the observation of CMMs. First, we gradually increase the magnetic field in the $z$ direction, while the STM tip continues to probe the edge and surface of the thin FI. Once the bulk of the FI passes through the gap closing phase, the FI surface is expected to be gapped. If the transition is topological, the gapless states on the edges should be observed by STM. Furthermore, the CMMs can be gapped in small FIs due to the finite-size hybridization. The existence of CMMs can be further supported by measuring the hybridization gap, which exponentially decays in magnitude as the size of the FI increases~\cite{Marcus}. However, STM cannot distinguish CMMs and chiral electron states. Hence, a second step of measuring the transport of the FI is required. As discussed in \cite{KangWang}, when the FI is sandwiched between two Chern insulators, its conductance should be $e^2/2h$ in the presence of the single CMM. Thus, the gapless spectrum on the edge of the FI and the measurement of $e^2/2h$ conductance should corroborate the presence of the CMM. 

Last but not least, the chirality of the Majorana edge states is determined by the magnetization direction of the FI layer, which can be manipulated by a weak magnetic field. This enables a convenient way to investigate rich Majorana physics arising from the interaction between Majorana states in proximity to each other. Through patterning the magnetic islands on the surface of STI and implementing a local magnetic field to control the magnetization of each island,  an array of CMMs with different chiralities can be realized in a solid state setup as illustrated in Fig.~4. For example, in a configuration with alternating magnetizations as shown in the top row of Fig.~4, there are two CMMs with the same chirality at the boundary shared by two adjacent FI islands. The two CMMs behave cooperatively like a single chiral electron state.  This chiral electron state splits at the open edge into two Majorana states propagating in opposite directions. In another configuration like the one depicted in the bottom row of Fig. 4, the magnetizations of all islands are aligned in the same direction. In this case, at the shared boundaries there are two counterpropagating CMMs which can be hybridized and generate an energy gap. On the open edge, the Majorana edge states on each island can form collectively a spatially extended Majorana state through tunneling between adjacent islands. In other words, this array of FI islands with the same magnetization is equivalent to a single-domain island with a larger dimension. This controllability of edge state chirality by an external magnetic field creates exciting opportunities for studying the interaction effect of Majorana fermions and further utilizing CMMs in practical applications of topological computation. We leave the detailed material simulations and experimental measurements for the proposed chiral Majorana platform for future research.


\section{Acknowledgements} The authors are indebted to W. S. Cole, C.-X. Liu, J. D. Sau, D. S. Sanchez, and I. Belopolski for discussions. 
CKC is supported by Microsoft Q and LPS-MPO-CMTC. GB is supported by startups from University of Missouri, Columbia. Work at Princeton is primarily supported by U.S. DOE.


\newpage

\begin{figure}
\centering
\includegraphics[width=16cm]{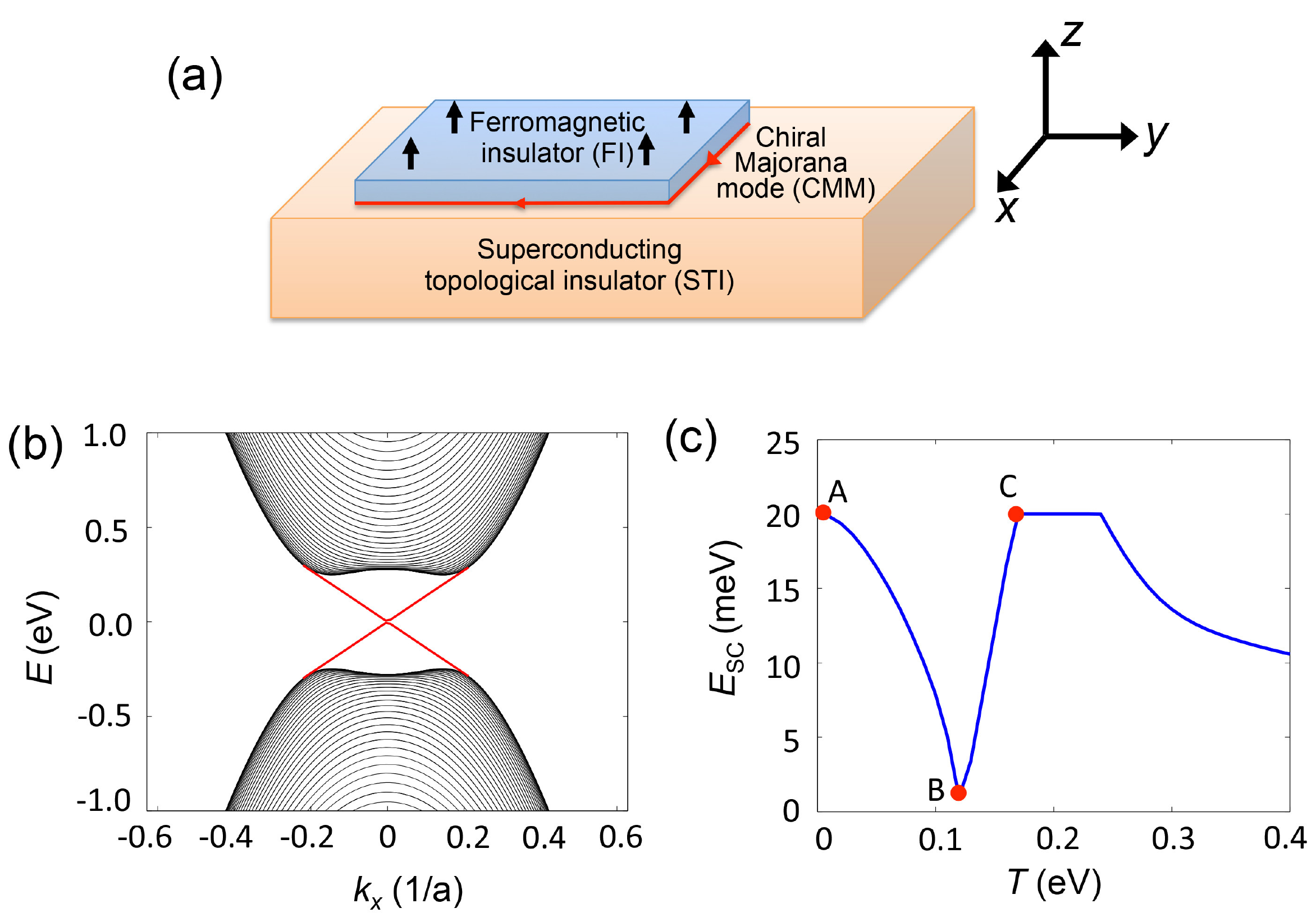}
\caption{(a) Schematic of a superconducting topological insulator with one surface covered by a ferromagnetic layer. (b) The tight-binding band structure of the topological insulator as $k_y=0$. The red line indicate the Dirac surface spectrum. (c) The superconducting energy gap as a function of the exchange parameter $T$. The superconducting order parameter $\Delta$ is set to be 20 meV and the chemical potential $\mu$ is 0.05 eV, which is within the TI bulk band gap. The details of the spectra at point A, B, and C are shown in \ref{fig2} (a), (b), and (c) respectively. 
}
\label{fig1}
\end{figure}

\newpage

\begin{figure}
\centering
\includegraphics[width=16cm]{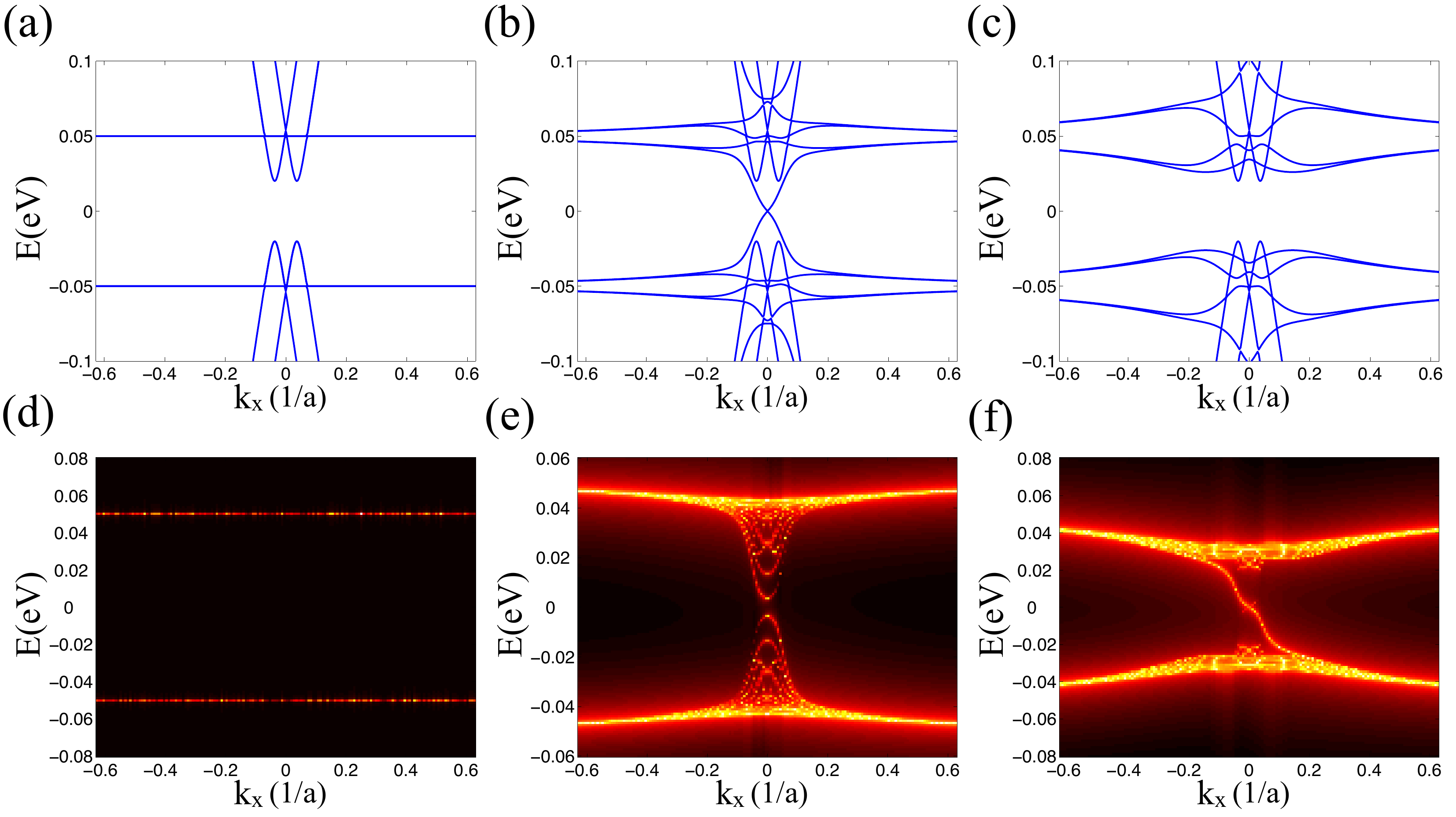}
\caption{The  band structure of the FI-STI system with $\mu$ = 0.05~eV, $\Delta$ = 20~meV, $M_z$ =~0.05 eV and different exchange coupling parameters (a) $T$ = 0, (b) $T$ = 0.122~eV, and (c) $T$ = 0.2~eV. The three exchange coupling parameters are marked in Fig.~1(c). In (a), the gapped Dirac cone is doubly degenerate for two surfaces of the STI film. In (b, c), the gapped Dirac cone in the band structure belongs to the bottom surface, which does not participate in hybridization with the ferromagnetic overlayer. On the top surface, the topological phase transition occurs in (b) and the gapped system has non-trivial topology in (c). (d-f) The corresponding projected edge spectrum for (a-c). The width of the ferromagnetic island is $L=100a$, where $a$ is the size of unit cell. The chiral Majorana mode appears after the topological quantum transition (bulk band touching). }
\label{fig2}
\end{figure}

\newpage

\begin{figure}
\centering
\includegraphics[width=16cm]{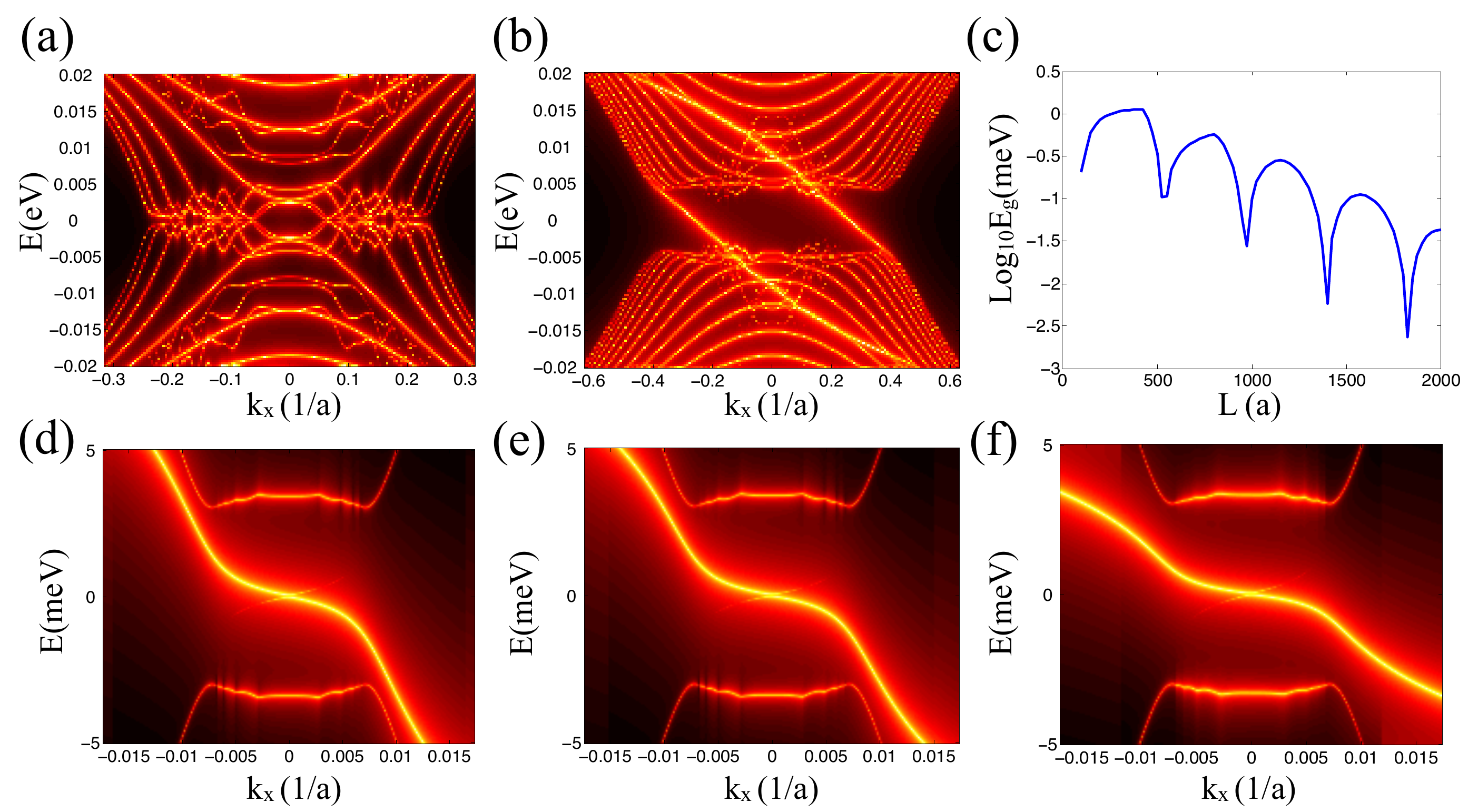}
\caption{
The projected spectra of the FI-STI heterostructure on one edge of FI with different physical parameters: (a,b) show the physics as the chemical potential is located in the conduction band $\mu=0.35$ eV, $L=60a$. (a) At $T=0.25$~eV, the interface between STI and FI exhibits multiple band touchings in moderate interval $T$. (b) At $T=0.35$~eV, two CMMs with the same chirality are present in the gap after the occurrence of multiple band touchings. (c-f) show the presence of the CMM for a \emph{small} superconducting order parameter $\Delta=3$~meV, $\mu=0.01$ eV, $T=0.25$ eV. The width of the FI ribbon is $L=2000a$. (c) The Majorana hybridization gap $E_{\rm{g}}$, which is the smallest energy scale for the entire system, exponentially decays with oscillation as the width ($L$) of the FI ribbon increases. $M_z=0.05$~eV. (d) A single CMM and $3$~meV superconducting gap are present in the projected edge spectrum. (e) $M_x=M_z=0.05/\sqrt{2}$~eV.  The edge spectrum is almost identical to the one without in-plane magnetic field as shown in (d). (f) $M_y=M_z=0.05/\sqrt{2}$~eV. The slope of the CMM is changed by the $y$-direction magnetic field.} 
\label{fig3}
\end{figure}
 
\newpage

\begin{figure}
\centering
\includegraphics[width=16cm]{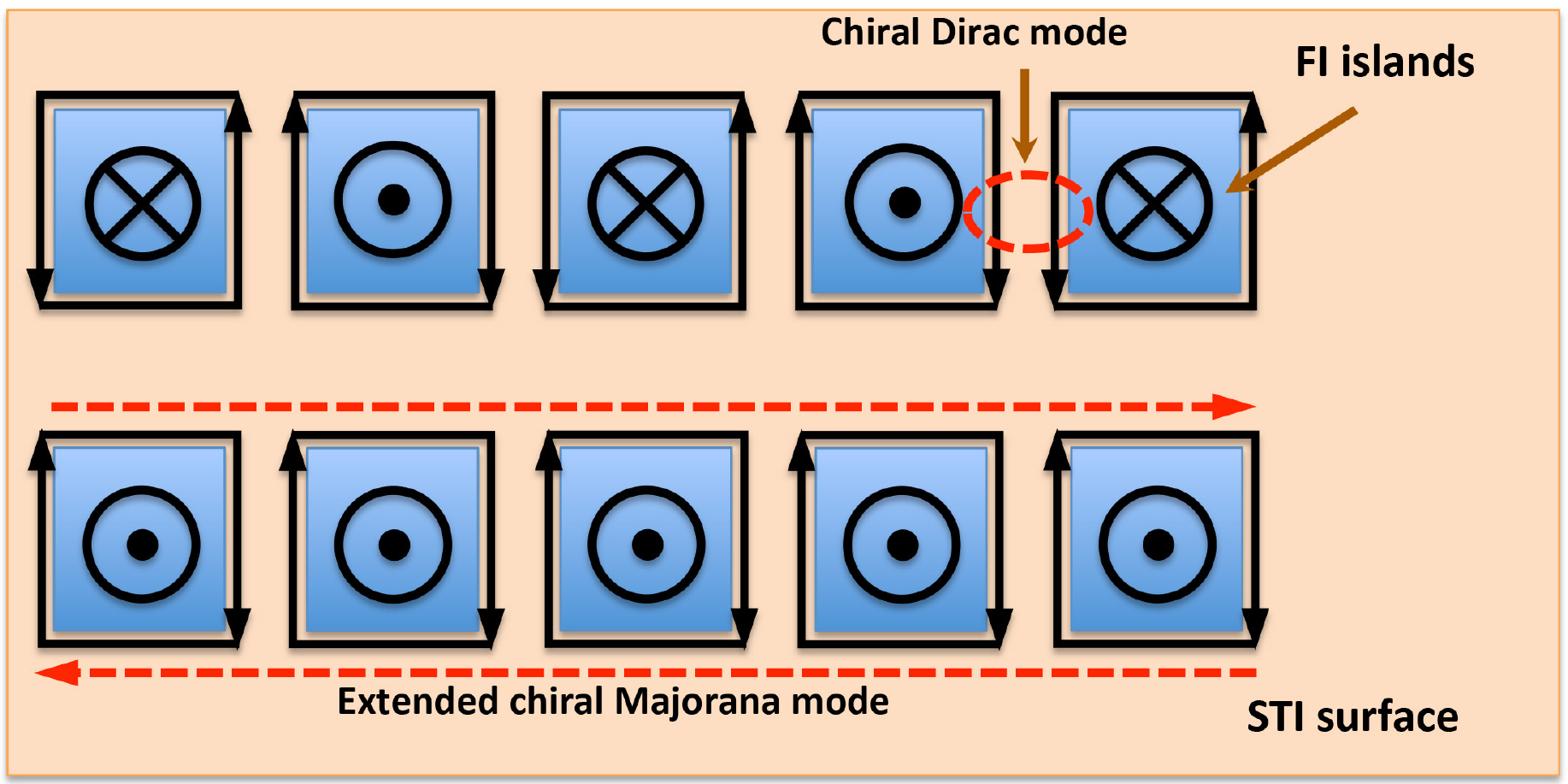}
\caption{
Schematic of a solid-state setup with multiple chiral Majorana modes. The orange areas represent the surface of a superconducting topological insulator and the blue areas ferromagnetic insulator islands. The magnetization of each island is indicated by the $\otimes$ and $\odot$ signs, which mean the directions into the screen and out of the screen, respectively. The propagation direction of the chiral Majorana states is given by the arrows.}
\label{fig4}
\end{figure}

\end{document}